\documentclass[12pt,fleqn,twoside]{article}
\usepackage{gc}

\input{epsf.tex}

\heads{A. Kleber, J. P. S. Lemos and V. T. Zanchin}
      {Thick shells and stars in Majumdar-Papapetrou general relativity}

\begin{document} 

 
\Title { Thick shells and stars in Majumdar-Papapetrou 
general relativity}

\Author{Antares Kleber\foom 1, Jos\'e P. S. Lemos\foom 2 and Vilson
 T. Zanchin\foom 3}
{\foom 1Observat\'orio Nacional - MCT,
  Rua General Jos\'e Cristino 77,
20921-400, Rio de Janeiro, Brazil, {anta@on.br}\\
\foom 2Centro Multidisciplinar de Astrof\'{\i}sica - CENTRA, 
 Departamento de F\'{\i}sica, Instituto Superior T\'ecnico,
Universidade T\'ecnica de Lisboa,
 Av. Rovisco Pais 1, 1049-001 Lisboa, Portugal, 
 {lemos@fisica.ist.utl.pt}\\
\foom 3Departamento de F\'{\i}sica, 
 Universidade Federal de Santa Maria, 
 97119-900 Santa Maria, RS, Brazil, zanchin@ccne.ufsm.br}


\vskip 1cm  
\Abstract
{The Majumdar-Papapetrou system is the subset of the
Einstein-Maxwell-charged dust matter theory, when the charge of each
particle is equal to its mass.  Solutions for this system are less
difficult to find, in general one does not need even to impose any
spatial symmetry a priori. For instance, any number of extreme
Reissner-Nordstr\"om solutions (which in vacuum reduce to extreme
Reissner-Nordstr\"om black holes) located at will is a solution. In
matter one can also find solutions with some ease. Here we find an
exact solution of the Majumdar-Papapetrou system, a spherically
symmetric charged thick shell, with mass $m$, outer radius $r_{\rm o}$, 
and inner radius $r_{\rm i}$.
This solution consists of three regions,
an inner Minkowski region, a middle region with extreme charged dust
matter, and an outer Reissner-Nordstr\"om region.  The matching of the
regions, obeying the usual junction conditions for boundary surfaces,
is continuous.  For vanishing inner radius, one obtains a Bonnor star,
whereas for vanishing thickness, one obtains an infinitesimally thin
shell.  For sufficiently high mass of the thick shell or sufficiently
small outer radius, it forms an extreme Reissner-Nordstr\"om
quasi-black hole, i.e., a star whose gravitational properties are
virtually indistinguishable from a true extreme black hole.  This
quasi-black hole has no hair and has a naked horizon, meaning that the
Riemann tensor at the horizon on an infalling probe diverges.  
At the critical value, when the mass is equal to the outer radius, 
$m=r_{\rm o}$, there is
no smooth manifold. Above the critical value when $m>r_{\rm o}$ 
there is no solution, the shell collapses into a singularity.
Systems with $m< r_{\rm o}$ are neutrally stable. Many of these
properties are similar to those of gravitational
monopoles.}

 


\vskip 1.7cm
\section{Introduction} 

One can couple Newtonian gravitation to Coulomb electric fields
without much effort. A particle with mass $m_1$ and charge $q_1$ is
certainly a solution of the system. If $q_1=m_1$, then one can put a
second particle with $q_2=m_2$, and $q_1$ and $q_2$ having the same
sign, anywhere. This is also a solution since the Coulomb repulsion
compensates exactly for Newtonian attractive force. And of course, one
can then put any number of particles, such that for each one the mass
equals the charge and the charges have all the same sign.  One can go
further, distribute the particles continuously to make a fluid, such
that the energy density $\rho$ and the charge density $\rho_{\rm e}$
of the charged fluid obey $\rho_{\rm e}=\rho$, and find this is going
to be a solution of the Newtonian-Coulomb gravitation.  There are
other solutions for the Newtonian-Coulomb gravitation, which do not
have $\rho_{\rm e}=\rho$, but in these cases the density profile is
singular somewhere, or one has to add some pressure to the fluid.

The general relativistic version of the  Newtonian-Coulomb gravitation 
is the Einstein-Maxwell system which admits a vacuum spherically symmetric, 
particle like solution, given by the Reissner-Nordstr\"om metric
\cite{reissner,nordstrom}. For pure vacuum this solution represents a
charged black hole when the mass is greater than the charge, an
extreme charged black hole when the mass is equal to the charge, and a
naked singularity when the mass is smaller than the charge.  Weyl in 1917
also had the idea of studying vacuum general relativity and
electromagnetism together \cite{weyl}.  Upon further imposing axial
symmetry he showed that if there is a functional relationship between the 
$g_{tt}$ metric component and the electric potential $\varphi$, that
relation should be of the form $g_{tt}=1+B\varphi+\varphi^2$.  In 1947
Majumdar \cite{majumdar} showed further that this relation holds for
no spatial symmetry at all, axial symmetry being a particular case.
Moreover, he showed that if $g_{tt}=(1+C\,\varphi)^2$, i.e., the above
relation is a perfect square, then the spatial metric is conformal to
the flat metric, the equations of the system reduce to a master
equation of the Poisson type, and the system needs not to be a vacuum
system, one can have charge matter with the charge density $\rho_{\rm
e}$ equal to the energy density $\rho$, $\rho_{\rm
e}=\rho$, thus finding the general relativity counterparts to the 
Newtonian-Coulomb gravitational solutions. 
Papapetrou explored the same set of ideas, his paper was
submitted one month before that of Majumdar \cite{papapetrou}.  In
vacuum, $\rho_{\rm e}=\rho=0$, the Poisson type equation reduces to
the Laplace equation, and any number of extreme Reissner-Nordstr\"om
black holes located at will is a solution, as was elucidated by Hartle
and Hawking \cite{hartlehawking}.  For non vacuum systems, systems
with $\rho_{\rm e}=\rho$, the Poisson equation to be solved is more
complicated and one usually imposes additional symmetries.  Along this 
line, new
solutions were found and discussed by Das \cite{das} and several other
authors \cite{cohen}-\cite{ivanov}.
Bonnor and collaborators, in a series of papers, have 
further developed the field, by constructing spherical 
symmetric Majumdar-Papapetrou charged stars 
\cite{bonnor1}-\cite{bonnor3}, 
as well as stars of other shapes, ellipsoidal for instance 
\cite{bonnor4}. Further developments appeared in Lemos and Weinberg 
\cite{lemosweinberg} where the causal structure and physical 
properties of new Majumdar-Papapetrou stars were analyzed. 
All these solutions may have some astrophysical appeal, 
since extreme dust matter can be realized in 
nature with slightly ionized dust 
(for instance, a dust particle with $10^{18}$ neutrons and 
1 proton, or an equivalent configuration, 
realizes extreme dust matter). The equilibrium needed 
for the dust to be ionized, 
such that $\rho_{\rm e}=\rho$ always, is certainly precarious 
but possible. 
Note further that the Majumdar-Papapetrou solutions, both in vacuum
and in matter, are also of great importance in supergravity and
superstring theories since, when embedded in these extended gravities,
they are supersymmetric \cite{gibbonshull,tod}, and saturate a bound
on the charge, the charge is equal to the mass, being then called BPS
(Bogomolnyi-Prasad-Sommerfield) objects.  For instance, low-energy
superstring theory admits a series of Majumdar-Papapetrou BPS type
solutions in higher dimensions, extreme black extended objects, which
can be constructed from extreme black holes by arraying in specific
directions (see, e.g, \cite{peet}), a process similar to finding
charged dust solutions.

Drawing on the work of Bonnor and collaborators 
\cite{bonnor2a}-\cite{bonnor2c} and on the 
techniques developed in \cite{lemosweinberg}, 
we report here on a new spherical symmetric 
Majumdar-Papapetrou star, a thick shell, 
made of three regions: 
an interior flat part described by the Minkowski 
metric up to radius $r_{\rm i}$, a middle 
matter part, from $r_{\rm i}$ to $r_{\rm o}$, 
described by extreme matter 
$\rho=\rho_{\rm e}$ with an appropriate metric, and 
an exterior region, from $r_{\rm o}$ to infinity, 
described by the extreme Reissner-Nordstr\"om solution.
The appearance of the three regions is novel. In section 
II we present the Majumdar-Papapetrou system in 
isotropic coordinates, which are the ones appropriate 
to find the solution. In section III we present the 
solution in Schwarzschild coordinates to 
further study the properties of the system. 
In section IV we conclude.

\section{The thick shell 
solution in isotropic coordinates: Equations and solution} 

\subsection{Equations and fields}

We will study Einstein-Maxwell theory coupled 
to charged dust (charged matter with zero 
pressure). 
The equation for the  Einstein-Maxwell-charged 
dust system is given by $(G=c=1)$
\begin{equation}
G_{ab}=8\pi \left( T_{ab}^{\rm dust}
+T_{ab}^{\rm em}\right)\,,
\label{einsteinmaxwellchargeddust}
\end{equation}
where $G_{ab}$ is the Einstein tensor. 
$T_{ab}^{\rm dust}$ is the dust part of 
the stress-energy tensor
\begin{equation}
T_{ab}^{\rm dust}=\rho \, u_au_b\,,
\label{mattertensor}
\end{equation}
with $\rho$ being the energy-density, and 
$u_a$ is the 
four-velocity of the fluid. 
$T_{ab}^{\rm em}$ is the electromagnetic part of 
the stress-energy tensor
\begin{equation}
T_{ab}^{\rm em}=
\frac{1}{4\pi}\left(
F_{a}^{c}F_{bc}-\frac14g_{ab}F^{cd}F_{cd}
\right)\,,
\label{emtensor}
\end{equation}
with $F_{ab}$ being the Maxwell tensor. 
The two Maxwell equations, are
\begin{equation}
{F_a^b}_{;b}=4\,\pi\,j_a
\,,
\label{maxwellequation1}
\end{equation}
\begin{equation}
F_{[ab;c]}=0
\,.
\label{maxwellequation2}
\end{equation}
where $j_a=\rho_{\rm e}\, u_a$ is the 
electric four-current, $\rho_{\rm e}$ is the electric 
charge density 
of the dust, and $[\,]$ denotes anti-symmetrization. 
Equation (\ref{maxwellequation2}) permits 
to define a vector potential $A_a$ such that 
\begin{equation}
F_{ab}=A_{b,a}-A_{a,b}\,.
\label{maxwelltensor}
\end{equation}
Now, for a static  pure electric system  one can make 
the choice
\begin{equation}
u_a=\frac{\delta_a^0\,}{U}\;,\quad A_a=\delta_a^0\varphi\,,
\label{staticelectricchoice}
\end{equation}
where $U$ and $\varphi$ are functions of the spatial 
coordinates, $U$ being identified with the gravitational 
potential, and $\varphi$ with the electric potential. 
Furthermore, it was then shown by Majumdar \cite{majumdar} 
in a very elegant paper, that in the special case of extreme 
dust matter, i.e.,  when the 
energy density is equal to the charge density, 
\begin{equation}
\rho_{\rm e}=\rho\,,
\label{majundarconditions}
\end{equation}
the metric can be put in form
\begin{equation}
ds^2=-\frac{dt^2}{U^2}+ U^2\left(dx^2+dy^2+dz^2\right)
\,,
\label{isotropicmetric}
\end{equation}
where $(t,x,y,z)$ are called isotropic coordinates. 
The Einstein-Maxwell-extreme dust matter system 
(\ref{einsteinmaxwellchargeddust}) then can be reduced to the 
following system of equations, 
\begin{equation}
\nabla^2 U=-4\pi\,\rho\,U^3
\,,
\label{equationforUandrho}
\end{equation}
\begin{equation}
\varphi=1-\frac{1}{U}
\,, 
\label{equationforphi}
\end{equation}
with $\nabla^2 \,\equiv\,\frac{\partial^2}{\partial x^2}
+\frac{\partial^2}{\partial y^2}+\frac{\partial^2}{\partial z^2}$
is the flat Laplacian operator. 
We should comment that there is a possible choice for the sign of 
the charge density, strictly speaking 
$\rho_{\rm e}=\pm\rho$. Above we have chosen the plus sign, had we 
chose the minus sign the sign of the electric 
potential also changes, i.e.,  $\varphi=1+1/U$. 
In order to not carry a $\pm$ label we stick to the plus sign of 
the charge density, knowing the results will not be altered. 

Thus, from equation (\ref{equationforUandrho}) one sees that 
solutions with no spatial symmetry at all, are solutions of 
this system. This is because particles with charge equal to mass,  
or matter systems with $\rho_{\rm e}=\rho$, 
exert zero force upon each other, or within themselves, 
a result valid in both 
Newtonian gravitation and general relativity, so that 
they can be distributed at will. Solutions 
of this Einstein-Maxwell-extreme dust matter system are generically 
called Majumdar-Papapetrou solutions \cite{majumdar,papapetrou}, 
a particular instance of these are the Bonnor stars 
\cite{bonnor1}-\cite{bonnor4}.

\subsection{Thick shell solution with three regions: 
interior  flat, middle charged dust, exterior 
extreme Reissner-Nordstr\"om}

Here we want to find a spherical symmetric solution 
in which case the
line element (\ref{isotropicmetric}) should be written as
\begin{equation}
ds^2=-\frac{dt^2}{U^2}+ U^2\left[dR^2+R^2\,
(d\theta^2+\sin^2\theta\,d\phi^2)
\right]
\,,
\label{isotropicmetricspherical}
\end{equation}
where $U=U(R)$, 
and the three-dimensional Laplacian is now $\nabla^2U=
\frac{1}{R^2} \frac{\partial}{\partial R}
\left( R^2 \frac{\partial U}{\partial R}\right)$.
The field equation (\ref{equationforUandrho}) 
can be solved by guessing a potential $U$, and then 
finding $\rho$. The solution is then complete 
because $\rho_{\rm e}$ and $\varphi$ follow directly. 
If the solution is physically acceptable, then it is 
of interest. One such spherical symmetric 
solution has been found by 
Bonnor and Wickramasuriya \cite{bonnor2a}-\cite{bonnor2c}, 
describing a matter configuration from $R=0$ to 
$R=R_{\rm o}$, and then vacuum outside, given 
by the extreme Reissner-Nordstr\"om metric. 
We find here a non trivial extension of this Bonnor star solution 
by adding a region internal to the matter that is flat, described 
by the Minkowski metric. We display for the first time a global 
Majumdar-Papapetrou exact solution that is composed of 
three regions, the interior vacuum 
Minkowski, the middle matter region, and the exterior extreme 
Reissner-Nordstr\"om electrovacuum region.

The system is characterized by its ADM mass $m$, 
and its inner and outer isotropic 
radii $R_{\rm i}$ and $R_{\rm o}$, respectively. 
The solution for $U(r)$ (and thus for the system) 
is 
\begin{equation}
U_{\rm flat}= 1+\frac{m}{R_{\rm o}}\left(1
+\frac{R_{\rm o}-R_{\rm i}}{2R_{\rm o}}    
\right) \,,\quad 0\leq R\leq R_{\rm i}\,,
\label{flatregionmetric}
\end{equation}
\begin{equation}
U_{\rm matter}= 1+\frac{m}{R_{\rm o}}\left[1
+ \frac{\left(R_{\rm o}-R_{\rm i}\right)^2-
\left(R-R_{\rm i}\right)^2}
{2R_{\rm o}\left(R_{\rm o}-R_{\rm i}\right)}
\right] \,,\quad R_{\rm i}\leq R\leq R_{\rm 0}\,,
\label{matterregionmetric}
\end{equation}
\begin{equation}
U_{\rm RN}= 1+\frac{m}{R}\,,\quad 
R_{\rm o}\leq R\,.
\label{RNregionmetric}
\end{equation}
With $U$ we can find through Equation 
(\ref{isotropicmetricspherical}) the metric 
for the whole spacetime, and through Equation 
(\ref{equationforphi}) the electric potential $\varphi$. 
The energy density is zero outside matter, 
and in the matter it is found using  
Equation (\ref{equationforUandrho}), to be 
\begin{equation}
\rho=\frac{3\,m}{4\,\pi\,R^3_{\rm 0}}
\frac{R_{\rm 0}}{R_{\rm 0}-R_{\rm i}}
\frac{1-\frac23\frac{R_{\rm i}}{R}}{U^3_{\rm matter}}
\,,\label{rho1}
\end{equation} 
with the charge density being then $\rho_{\rm e}=\rho$. 
Since $\rho\geq0$ throughout the spacetime, one can say 
that $U$ given in Equations 
(\ref{flatregionmetric})-(\ref{RNregionmetric}) yields so far  
a physically acceptable solution. 
Regular boundary conditions at the center and at 
infinity are obeyed, at the center the metric and 
matter fields are regular, and at infinity spacetime 
is asymptotically flat, the electric field decays as 
a Coulomb field, and the extreme dust matter density is 
zero. Moreover, 
since there are two interfaces one has to check 
whether the junction conditions 
are obeyed at these boundaries.  Indeed, it is found
that they are obeyed, with no need for thin shells, 
see Section \ref{thethickshell} for more details.
If one puts $R_{\rm i}=0$ one gets the Bonnor 
solution \cite{bonnor2b} with two regions, the matter 
region and the exterior extreme Reissner-Nordstr\"om region. 
Note also that on using (\ref{flatregionmetric}) on
(\ref{isotropicmetricspherical}) one clearly gets the Minkowski 
metric, but not in the usual form. To do so 
one has to define a $\bar t$ such that 
${\bar t}=t/U_{\rm flat}$, and an $\bar R$ such that 
${\bar R}=U_{\rm flat}\,R$. The coordinates in the 
matter and electrovacuum regions would change accordingly.
We will not do it here, since it is not necessary and 
gets cumbersome. 

The use of isotropic coordinates is good for finding 
solutions, as Majumdar has shown \cite{majumdar}, but 
the interpretation of the solution is 
best done in Schwarzschild coordinates, where 
the radial coordinate $r$ yields a  measure of 
circumferences, or of areas, 
of the type we are used to in flat space.

\section{The thick shell solution in Schwarzschild coordinates: 
Physical features}

\subsection{Equations and fields}
In Schwarzschild coordinates we write the metric in the form
\begin{equation}
ds^2=-B(r)\,dt^2+A(r)\,dr^2+
r^2\left(d\theta^2+\sin^2\theta\,d\phi^2\right)
\,,
\label{metricEMSch}
\end{equation}
and the electric Maxwell field is written as  
\begin{equation}
A_0=\varphi(r)\,,\quad A_i=0\,.
\label{MfieldEMspherical}
\end{equation}
One can use Equations 
(\ref{einsteinmaxwellchargeddust})-(\ref{maxwelltensor})
to find three non-trivial equations, 
\begin{equation}
\frac{(AB)'}{AB}=8\pi\,r\,\rho\,A
\,,
\label{equationforB-EM}
\end{equation}
\begin{equation}
\left[r\left(1-\frac{1}{A}\right)\right]'
=8\pi\,r^2\,\rho+\frac{r^2}{A\,B}\,\varphi'^2
\,,
\label{equationforA-EM}
\end{equation}
\begin{equation}
\frac{1}{r^2\sqrt{A\,}}
\left[\frac{r^2}{\sqrt{AB}}\,\varphi'\right]'
=-4\pi\rho_{\rm e}
\,.
\label{equationfophiEM}
\end{equation}
Upon using the Majumdar-Papapetrou condition 
$\rho_{\rm e}=\rho$ one should be able to find 
the thick shell solution in Schwarzschild coordinates. 
Of course, it is much easier to make  a simple coordinate 
transformation from the solution in isotropic spherical coordinates 
$(t,R,\theta,\phi)$ to the solution in Schwarzschild coordinates 
$(t,r,\theta,\phi)$. This we will do next.

\subsection{The thick shell solution}\label{thethickshell}

In Schwarzschild coordinates, where the  
the metric takes the form (\ref{metricEMSch}), 
one sees that the coordinate $r$ has some physical meaning, the 
area of a sphere of constant $r$ is given 
by $4\,\pi\,r^2$. 
By inspection of Equations (\ref{isotropicmetricspherical}) 
and (\ref{metricEMSch}) 
one finds that the relation between $r$ and $R$ is 
\begin{equation}
r=U\,R\,.
\label{relationofradialcoordinate}
\end{equation}
Then, one gets $B(r)=1/U^2$. From 
(\ref{relationofradialcoordinate}) one finds 
$dr/[1+\frac{R}{U}\frac{dU}{dR}(r)]
=UdR$. Thus $A(r)=1/[1+\frac{R}{U}\frac{dU}{dR}(r)]^2$. 
Note that from (\ref{relationofradialcoordinate}) one gets 
$R$ as a function of $r$, in general implicitly, but in some 
cases, such as here, it can be solved explicitly. 
Now we have to transform coordinates in each region. 
We begin with the easiest ones, the exterior and interior 
vacua regions, and then do the matter region. 
The three regions 
are schematically represented in Figure 1.

The exterior 
Reissner-Nordstr\"om region is trivial, and we will do 
it first.  From Equations 
(\ref{RNregionmetric}) and (\ref{relationofradialcoordinate}) 
one gets, 
for the whole Reissner-Nordstr\"om region 
\begin{equation}
R=r-m\,, 
\label{RasfunctionofrinRNregin}
\end{equation}
and in particular at the boundary, one has 
\begin{equation}
R_{\rm o}=r_{\rm o}-m
\,.
\label{RexternalasfunctionofrexternalinRNregin}
\end{equation}

\vskip 0.3cm 
\centerline{\epsffile{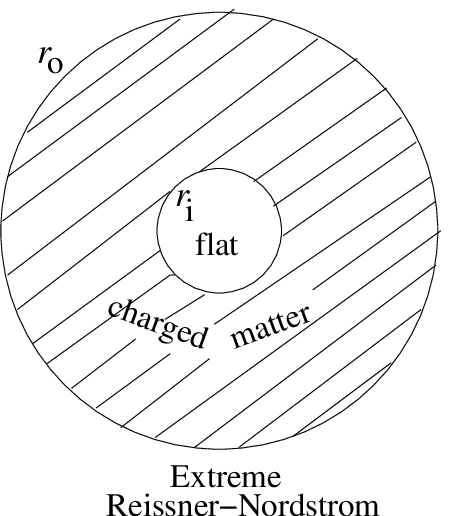}} 
\vskip 0.6cm 
{\noindent {\small Figure 1 - 
A schematic drawing of the three regions of the 
thick shell solution: the interior flat region, 
the middle charged matter region, and the exterior 
extreme Reissner-Nordstr\"om region.} } 
\vskip 0.3cm 

For the interior flat region one finds from  Equations 
(\ref{flatregionmetric}) and (\ref{relationofradialcoordinate})  
that $R_{\rm i}$ is 
a function of $r_{\rm i}$ and $r_{\rm o}$, i.e., 
$R_{\rm i}= R_{\rm i}(r_{\rm i},r_{\rm o})$, and 
we define  $f_{\rm i}\equiv R_{\rm i}(r_{\rm i},r_{\rm o})$. 
Then
\begin{equation}
f_{\rm i}=\left[
\frac{r_{\rm o}}{m}+\frac12
-\sqrt{\left(\frac{r_{\rm o}}{m}+\frac12\right)^2-
\frac{2\,r_{\rm i}}{m}}\;\,
\right]
\left(\frac{r_{\rm o}}{m}-1\right)\,m
\,. 
\label{Rinternalasfunctionofrinternalinflatregion}
\end{equation}
Then again from Equations 
(\ref{flatregionmetric}) and (\ref{relationofradialcoordinate})  
one has for the whole flat region
\begin{equation}
R=\frac{r}{
1+\frac{m}{r_{\rm o}-m}
\left[
1+\frac{r_{\rm o}-m-f_{\rm i}}{2(r_{\rm o}-m)}
\right]}
\,.
\label{Rasfunctionofrinflatregion}
\end{equation}

For the matter region, from Equations (\ref{matterregionmetric})
and (\ref{relationofradialcoordinate}), one has to solve a nasty cubic
equation for $R$, giving $R$ as a function of $r$. We denote this
function by $f(r)$, i.e., in the matter region, 
\begin{equation}
R= f(r)
\,,
\label{Rasfunctionofrinmatterregion}
\end{equation}
and give $f(r)$ in the Appendix A. Note that $f(r_{\rm i})=f_{\rm i}$. 

We can now write the metric in Schwarzschild coordinates 
for $0\leq r<\infty$. 
For the interior flat region is, 
\begin{equation}
ds^2=-\left(\frac{f_{\rm i}}{r_{\rm i}}\right)^2\,dt^2
+dr^2+r^2 \left(d\theta^2+\sin^2\theta\,d\phi^2\right)
\,,\quad 0\leq r\leq r_{\rm i}
\,. 
\label{metricflatSchw}
\end{equation}
For the matter region is 
\begin{eqnarray}
&ds^2=-\left(\frac{f(r)}{r}\right)^2\,dt^2
+\frac{dr^2}{\left(1-\frac{f^2(r)}{r}\,
\frac{m\,\left[f(r)-f_{\rm i}\right]}{(r_{\rm o}-m)^2
\,(r_{\rm o}-m-f_{\rm i})}\right)^2}\nonumber \\&
+\, r^2 \left(d\theta^2+\sin^2\theta\,d\phi^2\right)
\,,\quad  r_{\rm i}\leq r\leq r_{\rm o}
\,. 
\label{metricmatterSchw}
\end{eqnarray}
For the exterior Reissner-Nordstr\"om region one has 
\begin{eqnarray}
ds^2=-\left(1-\frac{m}{r}\right)^2\,dt^2
+\frac{dr^2}{(1-\frac{m}{r})^2}+
r^2 \left(d\theta^2+\sin^2\theta\,d\phi^2\right)
\,,\quad   r_{\rm o}\leq r
\,. 
\label{metricRNSchw}
\end{eqnarray}
The three regions join continuously, since the metric 
and the extrinsic curvature of the two boundaries 
match, as we show in Appendix B. .
A quick way to see this is to note that at the inner 
boundary the $g_{tt}$ part obviously matches, and 
the $g_{rr}$ part also matches since in the 
matter region one has $g_{rr}=1$ at $r=r_{\rm i}$. 
In the outer boundary one has $f(r_{\rm o})/r_{\rm o}=
1-m/r_{\rm o}$, so the $g_{tt}$ terms match, and as well 
the $g_{rr}$ are identical at $r_{\rm o}$.
One can also check
that $\left. \frac{\partial f(r)/r}{\partial r}\right)_{r_{\rm i}}=0$
so that the first derivatives of $g_{tt}$ match at the inner
boundary. As well, 
$\left. \frac{\partial f(r)/r}{\partial
r}\right)_{r_{\rm o}}= 2\left(1-\frac{m}{r_{\rm
o}^2}\right)\frac{m}{r_{\rm o}^2}$, so that the first derivatives of
$g_{tt}$ match at the outer boundary. The derivatives of $g_{rr}$ do
not match, but that is expected for a boundary surface. So $g_{rr}$ is
$C^0$ at both boundaries and the other metric functions are $C^1$ or
higher. This means that the Israel junction conditions are satisfied, 
yielding a boundary surface (see also Appendix B). With some extra effort 
one can smooth out the $g_{rr}$ component at the boundaries, 
so that $g_{rr}$ is $C^1$ or $C^2$, but we will not do it here.

The electric potential is taken from Equation 
(\ref{equationforphi}), $\varphi=1-\frac{1}{U}$, 
but now $U$ is seen as a 
function of $r$, 
(alternatively, $\varphi$ can be taken from $\varphi 
=1-\sqrt{|g_{tt}|}$ with the $|g_{tt}|$ for the 
three regions given in Equations 
(\ref{metricflatSchw})-(\ref{metricRNSchw})). 
Thus, we have 
\begin{equation}
\varphi=1-\frac{1}{1+\frac{m}{r_{\rm o}-m}\left(1
+\frac{r_{\rm o}-m-f_{\rm i}}{2(r_{\rm o}-m)}    
\right) } \,,\quad 0\leq r\leq r_{\rm i}\,,
\label{flatregionelectricpotential}
\end{equation}
\begin{equation}
\varphi=1-\frac{1}{U_{\rm matter}} \,,
\quad\quad\quad\quad\quad\quad 
\quad\quad
r_{\rm i}\leq r\leq r_{\rm 0}\,,
\label{matterregionelectricpotential}
\end{equation}
\begin{equation}
\varphi=1-\left(1-\frac{m}{r}\right)=\frac{m}{r}\,,
\quad\quad\quad\quad\quad\quad\quad
r_{\rm o}\leq r\,, 
\label{RNregionelectricpotential}
\end{equation}
where in Equation (\ref{matterregionelectricpotential}), 
$U_{\rm matter}(r)$ is now 
\begin{equation}
 U_{\rm matter}= 1+\frac{m}{(r_{\rm o}-m)}\left[1
+ \frac{\left(r_{\rm o}-m-f_{\rm i}\right)^2-
\left(f(r)-f_{\rm i}\right)^2}
{2(r_{\rm o}-m)\left(r_{\rm o}-m-f_{\rm i}\right)}
\right] \,,\quad r_{\rm i}\leq r\leq r_{\rm 0}\,,
\label{UmatterSchw}
\end{equation}
The electric potential is 
continuous and smooth at the boundaries.

The energy density can be also 
calculated in these coordinates. 
We have 
\begin{equation}
\rho =0 \,,
\quad\quad\quad\quad\quad\quad\quad
\quad\quad\quad\quad\quad\quad
\quad\quad\quad\quad\quad
 0\leq r < r_{\rm i}\,,
\label{flatregiondensity}
\end{equation}
\begin{equation}
\rho=\frac{3\,m}{4\,\pi(r_{\rm o}-m)^2\,
\left(r_{\rm o}-m-f_{\rm i}\right)}
\,
\frac{1-\frac{2f_{\rm i}}{3f(r)}}
{U_{\rm matter}(r)^3}
\quad\quad
r_{\rm i}\leq r\leq r_{\rm 0}\,,
\label{densitySchw}
\end{equation}
\begin{equation}
\rho =0 \,,
\quad\quad\quad\quad\quad\quad\quad
\quad\quad\quad\quad\quad\quad\quad\quad
\quad\quad\quad
r_{\rm 0} < r\,,
\label{RNregiondensity}
\end{equation}
where $U_{\rm matter}(r)$ is given in 
(\ref{UmatterSchw}). The energy density is a step function 
at $r_{\rm i}$ and $r_{\rm o}$, but this is no problem, 
jumps in the energy density are allowed.
It is easy to check that the energy density $\rho$ is everywhere positive
in the thick shell, as it should for a realistic fluid 
description.
The charge density 
follows from $\rho_{\rm e}=\rho$. 

Again, regular boundary conditions at the center and at 
infinity are obeyed, at the center the metric and 
matter fields are regular, and at infinity spacetime 
is asymptotically flat, the electric field decays as 
a Coulomb field, and the extreme dust matter density is 
zero. Moreover, as we have just showed the junction conditions 
at the outer and inner boundaries are also obeyed.

\subsection{Physical properties of the thick shell}

\subsubsection{From low gravitation configurations 
to extreme quasi-black holes}

The thick shell solution is given by the 
equations of the last subsection 
(\ref{metricflatSchw})-(\ref{RNregiondensity}). 
The solutions are characterized by the mass $m$, 
the outer and the internal radius, $r_{\rm o}$ 
and $r_{\rm i}$. An important dimensionless 
quantity is the ratio of the mass to outer radius, 
\begin{equation}
a=\frac{m}{r_{\rm o}}\,.
\label{parameter-a}
\end{equation}
\noindent This quantity is a measure of how general relativistic 
is the system. Small $a$ means the thick shell is very dispersed, 
Newtonian gravitation might suffice. For large $a$ 
general relativistic effects come into play and eventually 
at $a\rightarrow a_{\rm crit}=1$  ($m\rightarrow r_{\rm o}$) 
an event horizon is about to form. This is displayed 
in Fig. 2, where we draw  the dependence of the metric, electric 
and  density potentials as a function of radius, for 
$a$ small and $a_{\rm crit}$. 

The function $A$ signals the appearance 
of an event horizon, when $1/A$  passes through a zero, a horizon is formed, 
in this case it should be an extreme horizon, since $1/A$ gets a double 
zero (it is not strictly a double zero, since it is not smooth at 
$r_{\rm o}$, but with some care, either analytically or 
numerically one can smooth it out). 
The function $B$ is the redshift function, with $B=0$ meaning infinite
redshift. It is seen from equation
(\ref{Rinternalasfunctionofrinternalinflatregion}) that
for $a_{\rm crit}$ the parameter $f_{\rm i}$ is identically zero, which
means $B=0$ in the region $0\leq r\leq r_{\rm i}$. This is true also in the
region $r_{\rm i}< r \leq r_{\rm o}$, as can be checked from equations
(\ref{solution}) and (\ref{approximatesolution}) for the function
$f(r)$ in the Appendix A. Hence, at $a_{\rm crit}$ the redshift is
infinite at the horizon and in the whole region inside.  This means
that in fact a true black hole does not form, since inside there is no
smooth manifold. One refers to the location 
of the minimum in the function $1/A$, when $a\rightarrow a_{\rm crit}=1$, 
as the quasi-horizon, and the object is a quasi-black hole
\cite{lemosweinberg}. 
The product $(AB)^{1/2}$ is an important
function, it gives whether the horizon is naked or not, see the last 
two paragraphs of this section for a more specific discussion on this point. 
The electric potential $\varphi$ 
shows that at $a_{\rm crit}$, the electric potential 
outside the horizon is Coulombian, the black hole has 
no hair. 
The density field, shows a very interesting feature, 
when the star is about to form a quasi-black hole, there is still 
equilibrium, nothing special happens, its surface 
being a quasi-horizon with radius $r_*$ arbitrarily 
closed to the extreme Reissner-Nordstr\"om black hole 
radius $r_{\rm h}$.
Note that the figures displayed in Figure 2 correspond to Figures 
2-6 of 
\cite{lemosweinberg}. The curves for $A$, $B$,
$(AB)^{1/2}$ and $\phi$ show similar behaviors in both
cases, since for $r>r_{\rm o}$ they are the same functions and
for $r<r_{\rm o}$ they are power law functions. 
\vskip 0.3cm 
\centerline{\epsffile{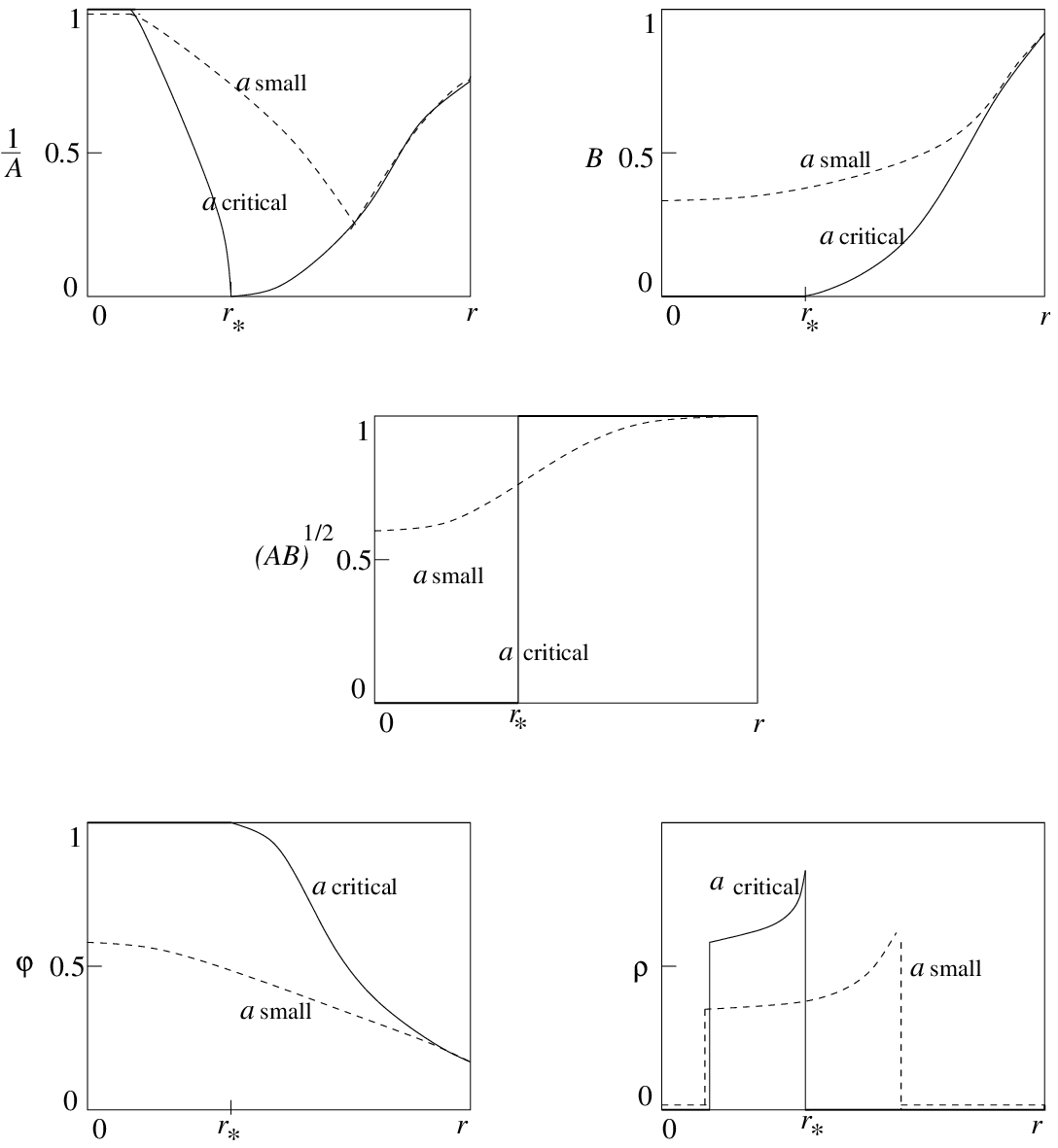}} 
\vskip 1mm 
{\noindent {\small Figure 2 - 
Behaviour of the metric, electric and matter functions in terms 
of the radius $r$. Note that $r_*$ is $r_{\rm o}$ critical.} } 
\vskip 0.3cm

There are two points worth noting in the meantime: 
(i) It is interesting that on one
hand these thick shell Majumdar-Papapetrou stars superseed the
Buchdall bounds for perfect fluid matter stars \cite{buchdall}, where
the radius of the star obeys $r_{\rm star}\geq \frac98 r_{\rm h}$,
with $r_{\rm h}=2m$, on the other hand, when they are about to form a
black hole the spacetime turns into a degenerate one, so that there is
no smooth manifold.  (ii) Although we do not discuss in detail in this
paper what happens above criticality, i.e., for $a>a_{\rm crit}$
($m>r_{\rm o}$), 
we note that in this case one would have a shell of matter inside
an extreme black hole at $r_{\rm h}=m$. Such a solution would be
everywhere free from curvature singularities and, following a theorem
by Borde \cite{borde}, this would mean that the topology of spacelike slices
in this black hole spacetime would change from a region where they
are noncompact to a region where they are compact, in the interior
(an example of this appears in the Bardeen model \cite{beato1,beato2} ).
In our case this in fact does not happen, there are no solutions with
$m>r_{\rm o}$, the shell collapses into a singularity.

We now dwell further on the nakedness of the extreme quasi-horizon. A
horizon is defined as 
naked when the tidal forces suffered by an infalling
particle at the horizon are infinite \cite{horowitz}. 
For a usual black hole spacetime, such as Schwarzschild or
Reissner-Nordstr\"om, 
the Riemann tensor $R_{abcd}$ is nonsingular at the horizon, and  
nothing special happens to an observer or a particle 
crossing it. But for some other black holes, such as some black 
holes with a dilaton field and the quasi-black hole 
studied here, the situation can be different. 
Due to the acceleration of the observer or particle, the 
Riemann tensor projected in a tetrad frame freely falling 
with it, may be divergently different from the Riemann 
tensor in the static coordinates. For the metric 
(\ref{metricEMSch}) one has that 
$R_{{\hat t}{\hat \theta}{\hat t}{\hat \theta}}$, where a hat 
means that the quantity is evaluated on a tetrad freely 
falling frame, 
is given by $R_{{\hat t}{\hat \theta}{\hat t}{\hat \theta}}
=-\frac{1}{2\,r}\,\frac{d\,}{dr}\left[\frac{E^2}{AB}-\frac{1}{A}\right]$, 
and similarly for  $R_{{\hat t}{\hat \phi}{\hat t}{\hat \phi}}$. 
The important
function here is the product $(AB)^{1/2}$. When $(AB)=1$, 
as is the case in the Schwarzschild or
Reissner-Nordstr\"om black holes, nothing special happens. But, 
when $(AB)^{1/2}\rightarrow0$, one has 
$R_{{\hat t}{\hat \theta}{\hat t}{\hat \theta}}\sim \frac{1}{AB}$
and thus this local component of the Riemann tensor at the 
horizon, i.e., the 
local component of the Riemann tensor as measured by a freely falling
observer passing through the horizon, diverges, and the horizon is naked. 
It is also interesting to relate this divergence to the proper 
time an observer or a particle take to make a return trip 
in the quasi-black hole spacetime. 
To see this, send a massive particle
from a large radius $r$, through the thick shell, so that it turns
around and comes back \cite{lueweinberg}. Due to the staticity and
spherical symmetry of the metric one can define a conserved energy and
a conserved angular momentum per unit mass for the particle,
$J=r^2\frac{d\phi}{d\tau}$, $E=B(r)\frac{d\,t}{d\tau}$.  Then from
Equation (\ref{metricEMSch}) one has
\begin{equation}
\frac{d\tau}{dr}=
\frac{(AB)^{1/2}}{\left[E^2-B\left(\frac{J^2}{r^2}+1\right)\right]^{1/2}}
\,.
\label{particlepropertime}
\end{equation}
We are interested in the quasi-horizon configuration, the one which  
is about to form a horizon but not quite. Call the radius 
of this configuration $r_*$. Call also 
$\left. \frac{1}{A}\right|_{r_*}=\epsilon$ 
where $\epsilon\rightarrow0$ at the critical solution. 
Then one has from Equation (\ref{particlepropertime}) that 
the particle spends a proper time inside the star of the 
order of 
$\Delta\,\tau\simeq \frac{r_*}{E}
(AB)^{1/2}|_{r_*}\sim \epsilon^{1/2}$ 
(indeed, put $r_{\rm i}=0$ to simplify (this does not alter the 
result), then for $1-\frac{r}{m}=\frac{\epsilon}{6}$ and 
$\frac{r_{\rm o}}{m}-1=\alpha\epsilon$, with $\alpha$ a number 
of order one and $\epsilon<<1$, one 
has $f(r)=(r_{\rm o}-m)(1-(\epsilon)^{1/2}/3)$, and thus 
$1/A^{1/2}=1-(m/r)[f(r)^3/(r_{\rm o}-m)^3]=\epsilon^{1/2}$, 
so $1/A=\epsilon$, and $B^{1/2}=f(r)/r=\alpha\epsilon$, so that 
$(AB)^{1/2}=\alpha\,\epsilon^{1/2}\;)$. 
For the quasi-horizon this time is extremely short, 
and at the critical configuration is zero, since 
$(AB)^{1/2}$ vanishes for $r<r_*$. 
In addition, one finds 
$R_{{\hat t}{\hat \theta}{\hat t}{\hat \theta}}
\sim \frac{1}{(\Delta\,\tau)^2}$. Since 
$\Delta\,\tau$ goes to zero at the critical solution, 
the local Riemann tensor 
diverges, as previously stated. 
Another interesting time, is the time $\Delta\,t$ the particle takes
in its trip for a coordinate observer, given by 
$\Delta\,t\sim \frac{r_*} {\epsilon^{1/2}}$.  At the quasi-critical solution
this time is arbitrarily large.  Thus, for an external observer the
particle takes an arbitrarily quasi infinite time to come back. The
observer cannot probe the inside, and associates it with an entropy
for the star (see \cite{lueweinberg} for a 
discussion on quasi-black hole entropy).

\subsubsection{Penrose diagrams}

One can draw the Penrose diagrams for the two 
configurations of interest,
the quasi-black hole with $a\rightarrow m/r_{\rm o}$ quasi-critical, and
the black hole with the same mass parameter, see Figure 3. For
the quasi-black hole, the Penrose diagram is the same as for $a$
small. In these cases, the spacetime is asymptotically Minkowskian,
and the origin of the coordinates $r=0$ and the surface of the thick
shell star $r_{\rm o}=r_*$ are timelike surfaces.  
On the other hand,
for the true black hole, spacetime is still asymptotically Minkowskian,
with $r_{\rm h}$ being a true horizon, 
whereas $r=0$ is now a curvature singularity.

Note that at criticality, $(m=r_{\rm o}$), there is no Penrose diagram
for the quasi-black hole. Since time has disappeared from spacetime,
the solution is now degenerate. Above criticality, $(m>r_{\rm o}$),
there are no static solutions.
\vskip 0.1cm 
\centerline{\epsffile{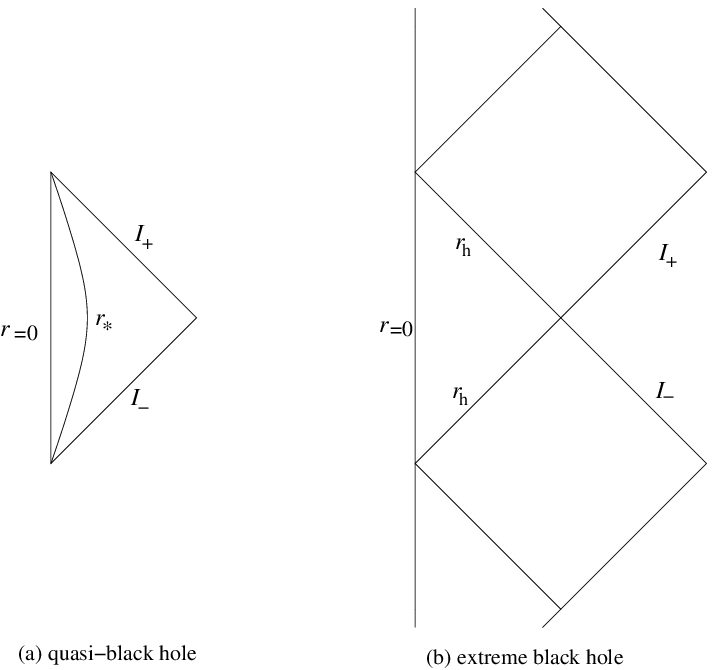}} 
\vskip 3mm 
{\noindent {\small Figure 3 - 
Penrose diagrams for (a) the quasi-black hole, and (b) the extreme 
black hole, are shown. } } 
\vskip 0.1cm

\subsection{Two  limits: stars and thin shells}

There are two interesting limits of the Majumdar-Papapetrou 
thick shell solution. 

\subsubsection{Bonnor stars}

If one takes the limit of a vanishing inner radius 
$r_{\rm i}=0$ one obtains a Bonnor star 
\cite{bonnor2a}-\cite{bonnor2c}. 
This solution has two regions, the interior from 
$r=0$ to $r=r_{\rm o}$, made 
of extreme charged matter, and the exterior, an 
extreme Reissner-Nordstr\"om region. The Bonnor star 
has the same properties of the thick shell solution, 
when the star is compact enough it forms an extreme 
Reissner-Nordstr\"om quasi-black hole with a naked horizon
(see \cite{lemosweinberg} for more details).

\subsubsection{Infinitesimally thin shells} 

Another important limit is the infinitesimally 
thin shell limit, when the thickness of the shell 
goes to zero, 
$r_{\rm i}\rightarrow r_{\rm o}$. The important 
quantity to keep track is the energy density 
$\rho$ (an energy per unit volume) 
that must go over into a surface energy
density $\sigma$ (an energy per unit area).  
From Equation (\ref{densitySchw}) one obtains,
when $r_{\rm o}\rightarrow r_{\rm i}$, 
$\rho=\frac{m}{4\pi\,r_{\rm o}^2}
\left(\frac{1-\frac{m}{r_{\rm o}}}
{r_{\rm o}-r_{\rm i}}\right)$, in first order 
in $r_{\rm o}-r_{\rm i}$. 
The proper surface energy density of the 
thin shell should be defined as $\sigma=
{\rm lim}_{r_{\rm o}\rightarrow r_{\rm i}}
\left(\frac{r_{\rm o}-r_{\rm i}}{1-\frac{m}{r_{\rm o}}} 
\right)\,\rho$, where the factor $(r_{\rm o}-r_{\rm i})$
is as in Newtonian theory, and the factor 
$1/(1-\frac{m}{r_{\rm o}})$ takes care of the proper 
length (see metric (\ref{metricEMSch})). Thus, we find 
\begin{equation}
\sigma=\frac{m}{4\,\pi\,r_{\rm o}^2}
\,, 
\label{surfacedensity}
\end{equation}
as it is expected for an infinitesimally thin shell. 
The three regions are now, an inner
flat region, a thin shell, and an exterior 
Reissner-Nordstr\"om vacuum. The metric is now 
\begin{equation}
ds^2=-\left(1-\frac{m}{r_{\rm o}}\right)^2\,dt^2
+dr^2+r^2 \left(d\theta^2+\sin^2\theta\,d\phi^2\right)
\,,\quad\quad 0\leq r < r_{\rm o}
\,, 
\label{metricflatSchwthin}
\end{equation}
\begin{equation}
ds^2=-\left(1-\frac{m}{r_{\rm o}}\right)^2dt^2
+\, r^2 \left(d\theta^2+\sin^2\theta\,d\phi^2\right)
\,,\quad\quad\quad\quad\quad\quad  
r= r_{\rm o}
\,,  
\label{metricmatterSchwthin}
\end{equation}
\begin{eqnarray}
ds^2=-\left(1-\frac{m}{r}\right)^2\,dt^2
+\frac{dr^2}{(1-\frac{m}{r})^2}+
r^2 \left(d\theta^2+\sin^2\theta\,d\phi^2\right)
\,,\quad   r_{\rm o} \leq r
\,.
\label{metricRNSchthinw}
\end{eqnarray}
Note that the $g_{rr}$ component has a step at 
$r_{\rm 0}$ from $1$ to 
$1/\left(1-\frac{m}{r_{\rm o}}\right)^2$, meaning 
that the junction conditions are not satisfied there. 
This is a minor problem, one can 
always see it as a very thin shell (not infinitesimally thin), 
where there is no discontinuity, although the slope 
of the $g_{rr}$ function is very high.  The electromagnetic field 
can be taken from $\varphi=1-\sqrt{|g_{tt}|}$ in equations 
(\ref{metricflatSchwthin})-(\ref{metricRNSchthinw}). 
The thin shell  
has analogous properties to the thick shell solution.  
When it is compact enough, for $a\rightarrow a_{\rm crit}=1$  
(i.e.,  $m\rightarrow r_{\rm o}$) it 
forms an extreme 
Reissner-Nordstr\"om quasi-black hole with a naked horizon. 
When the precise equality holds, $a= a_{\rm crit}$, 
the redshift function $B$ is zero 
not only at the horizon but also in the whole region inside, meaning 
that in fact a true black hole does not form, since inside there is 
no smooth manifold. This signals either a change in topology 
or instability.
Indeed, one can ask what happens for $a>a_{\rm crit}$, $m>r_{\rm o}$? 
In such a case a solution does not exist, the thin shell collapses into
a singularity. Note also that for the same mass parameter $m$
there is also the branch of pure extreme Reissner-Nordstr\"om 
black holes.

\section{Conclusions} 

We have found an exact solution of the Majumdar-Papapetrou system, 
the thick shell star solution, 
consisting of three regions, an inner Minkowski region, a 
middle region with extreme charged matter, and an outer 
Reissner-Nordstr\"om region. The system is neutrally stable, 
as all the systems of Majumdar-Papapetrou type are. 
For sufficiently high mass, or sufficiently small outer 
radius, at almost the critical value, 
the thick shell forms an extreme Reissner-Nordstr\"om quasi-black
hole, with no hair and with a naked horizon, i.e., the 
Riemann tensor at the horizon on an infalling probe diverges. 
At the critical value there is
no smooth manifold. Above the critical value when $m>r_{\rm o}$ 
one has an extreme shell inside an event horizon collapsing into a
singularity.
All these properties are similar to the properties found for 
gravitational monopoles when gravity takes care of the system 
\cite{lueweinberg}. In another work we will explore these 
similarities which are much too striking to pass without 
some attention \cite{lemosweinberg,lemos}.  

Thick shell solutions are rare in the literature. 
For a discussion of the generalization of 
the Israel thin shell junction conditions \cite{israel} 
to thick shells see \cite{mansouri}, and for an application 
to the covariant entropy bound in Tolman-Bondi spacetimes 
see \cite{gao}.

\Acknow
{We thank Erick Weinberg for conversations. JPSL thanks Columbia
University for hospitality, and acknowledges financial support from
the Portuguese Science Foundation FCT and FSE for support, through
POCTI along the III Quadro Comunitario de Apoio, reference number
SFRH/BSAB/327/2002, and through project PDCT/FP/50202/2003. 
VTZ thanks Observat\'orio Nacional do Rio de
Janeiro for hospitality.}

\newpage

\appendix

\section{The isotropic coordinate $R$ as a function of the 
Schwarzschild coordinate $r$ in the matter region}

We want to invert the equation
\begin{equation}
r=R\,U_{\rm matter}=
R\,\left[ 1+ \frac{m}{R_{\rm o}}
\left(1+
\frac{(R_{\rm o}-
R_{\rm i})^2-(R-R_{\rm i})^2}{2R_{\rm o}(R_{\rm o}-R_{\rm i})}
\right)
\right]
\,. 
\label{definingSchwr}
\end{equation}
In order to condense expressions, define 
\begin{equation}
a=-\frac{E}{F}-\frac13\,R_{\rm i}^2\,,\quad 
b=\frac{r}{F}+\left(-\frac{2\,E}{3\,F}+\frac{2R_{\rm i}^2}{27}
\right)\,R_{\rm i}
\,, 
\label{def1}
\end{equation}
with 
\begin{equation}
E=1+\frac{m}{R_{\rm o}}\left(
1+\frac{R_{\rm o}-R_{\rm i}}{2R_{\rm o}}
\right)\,,\quad 
F=\frac{m}
{2R_{\rm o}^2(R_{\rm o}-R_{\rm i})
}\,,
\label{def2}
\end{equation}
where $R_{\rm o}=r_{\rm o}-m$ as in 
(\ref{RexternalasfunctionofrexternalinRNregin}), 
and $R_{\rm i}=R_{\rm i}(r_{\rm i},r_{\rm o})$ 
as in (\ref{Rinternalasfunctionofrinternalinflatregion}).
Then through further definitions
\begin{equation}
A={}^{3}\sqrt{
-\frac{b}{2}+
\sqrt{\frac{b^2}{4}+\frac{a^3}{27}}}\,,\quad 
B={}^{3}\sqrt{
-\frac{b}{2}-
\sqrt{\frac{b^2}{4}+\frac{a^3}{27}}}
\,, 
\label{def3}
\end{equation}
the solution is $R=f(r)$, with 
\begin{equation}
f(r)=\frac23\,f_{\rm i}
-\frac{A+B}{2}-\frac{A-B}{2}\sqrt{-3}
\,. 
\label{solution}
\end{equation}
For $r$ and $r_{\rm i}$ small one finds
\begin{equation}
f(r)-f_{\rm i}=\frac{2(r_{\rm o}-m)}{2r_{\rm o}+m}
\left(r-r_{\rm i}\right)
\,,
\label{approximatesolution}
\end{equation}
where we have used
(\ref{RexternalasfunctionofrexternalinRNregin}) and 
(\ref{Rinternalasfunctionofrinternalinflatregion}).

\section{The matching conditions of the the thick 
shell}

We will follow Israel \cite{israel} to show that the 
two boundaries at $r_{\rm o}$ and $r_{\rm i}$ are 
boundary surfaces, i.e., do not need extraneous 
thin shells. 
We need to show that the metric, or first fundamental 
form, is continuous at the surface,  
$g_{ab}^+=g_{ab}^-$, and the extrinsic curvature, 
or second fundamental 
form, is also continuous,  $K_{ab}^+=K_{ab}^-$. 
For the extrinsic curvature, we note first 
that, since the surface we use is spherical 
the normal $n_a$ to the surface has only a radial component 
$n_r=\sqrt{g_{rr}}$. Then the extrinsic curvature has the 
form 
\begin{equation}
K_{ij}^{\pm}=-n_{r} \Gamma ^{r \pm}_{\;\;\alpha
\beta}\;\frac{\partial x^{\alpha}}{\partial \xi ^{i}} \,
\frac{\partial x^{\beta}}{\partial \xi ^{j}} 
\,,
\label{extrinsiccurvaturegeneral}
\end{equation}
where $\xi^i$ are the intrinsic coordinates of the surface. 
Now we analyze the junction at the outer and inner surfaces.

\begin{description}
\item[Outer boundary, $r=r_{\rm o}$:]

For the surface $r_{\rm o}$ we adopt the metric
\begin{equation}
ds^2=-\left(1-\frac{m}{r_{\rm o}}\right)^2\,d\tau^2+
r^2(d\theta^2+\sin^2 d\phi^2)
\,,
\label{metricoutersurface}
\end{equation}
with the intrinsic coordinates $\xi^i$ being 
$\xi^i=(\tau,\theta,\phi)$. Then 
one has $f(r_{\rm o})/r_{\rm o}=
1-m/r_{\rm o}$, so from Equations (\ref{metricmatterSchw})
and (\ref{metricRNSchw}) one sees that
the $g_{tt}^+$ and $g_{tt}^-$ 
match. The $g_{rr}^+$ and $g_{rr}^-$ also match, 
as well as the terms for the angular part.  
Now, at $r_{\rm o}$ coming for the interior 
one finds 
$K_{\tau\tau}^-=-n_r\,
\Gamma ^{r -}_{\;\;tt}\,
\frac{dt}{d\tau}\frac{dt}{d\tau}$.
By construction $\frac{dt}{d\tau}=1$. One also has 
$n_r=\frac{1}{1-\frac{f^2}{r}
\frac{m(f-f_{\rm i})}
{(r_{\rm o}-m)^2(r_{\rm o}-m-f_{\rm i})}}|_{r_{\rm o}}
=\frac{1}{1-\frac{m}{r_{\rm o}}}$. Noting 
that $\frac{df}{dr}|_{r_{\rm o}}=1$ one finds 
$\Gamma ^{r -}_{\;\;tt}=\frac{m(r_{\rm o}-m)^3}{r_{\rm o}^5}$. 
Thus, $K_{\tau\tau}^-=\frac{m(r_{\rm o}-m)^2}{r_{\rm o}^4}$.
The extrinsic curvature coming form the exterior 
Reissner-Nordstr\"om region can also be calculated, one 
finds  $K_{\tau\tau}^+=\frac{m(r_{\rm o}-m)^2}{r_{\rm o}^4}$.
So they match. One can also check that 
$K_{\theta\theta}$ and $K_{\phi\phi}$ also match.

\item[Inner boundary, $r=r_{\rm i}$:]
For the surface $r_{\rm i}$ we adopt the metric
\begin{equation}
ds^2=-\left(\frac{f_{\rm i}}{r_{\rm i}}\right)^2\,d\tau^2+
r^2(d\theta^2+\sin^2 d\phi^2)
\,.
\label{metricinnersurface}
\end{equation}
From Equations (\ref{metricflatSchw}) and (\ref{metricmatterSchw}) 
one sees that the $g_{tt}$ part matches, and 
the $g_{rr}$ part also matches since in the 
matter region one has $g_{rr}=1$ at $r=r_{\rm i}$. Likewise 
for $g_{\theta\theta}$ and $g_{\phi\phi}$. 
Now, at $r_{\rm i}$ coming for the matter 
one finds 
$K_{\tau\tau}^+=-n_r\,
\Gamma ^{r -}_{\;\;tt}\,
\frac{dt}{d\tau}\frac{dt}{d\tau}$. 
By construction $\frac{dt}{d\tau}=1$. One also has 
$n_r=1$. Noting 
that $\frac{df}{dr}|_{r_{\rm i}}=\frac{f}{r}$ one finds 
$\Gamma ^{r +}_{\;\;tt}=0$. 
Thus, $K_{\tau\tau}^+=0$.
The extrinsic curvature coming form the interior 
flat region can also be calculated, one 
finds  $K_{\tau\tau}^-=0$.
So they match. One can also check that 
$K_{\theta\theta}$ and $K_{\phi\phi}$ also match. 

\end{description}

\noindent So the metric and the extrinsic curvature are continuous 
as is required. The electric potential $\varphi$ 
and its first derivative at the boundaries are continuous 
as is also required.

\newpage


\end{document}